\documentclass[12pt]{article}
\usepackage{amsmath, amsthm, amssymb}
\usepackage{hyperref} 

\def\pdot{\dot{p}}
\def\qdot{\dot{q}}
\def\rdot{\dot{r}}
\def\adot{\dot{\a}}
\def\bdot{\dot{b}}
\def\zdot{\dot{z}}
\def\gdot{\dot{\g}}
\def\vdot{\dot{v}}
\def\Ldot{\dot{L}}
\def\rdot{\dot{r}}
\def\wbar{\bar{w}}
\def\mbar{\bar{m}}

\def\bolda{{\hbox{\bf a}}}
\def\boldg{{\hbox{\bf g}}}
\def\bolde{{\hbox{\bf e}}}
\def\cov{\hbox{\rm Cov}}
\def\zbar{\bar{z}}
\def\zdotbar{\dot{\bar{z}}}

\def\vdotbar{\dot{\bar{v}}}
\def\mdotbar{\dot{\bar{m}}}
\def\logdot{\dot{\log}}
\def\E{\hbox{E}}
\def\a{\alpha}
\def\b{\beta}
\def\d{\delta}
\def\e{\epsilon}
\def\g{\gamma}
\def\th{\theta}
\def\bar#1{\overline{#1}}
\def\ovr#1#2{{{#1}\over{#2}}}
\def\dovr#1#2{\ovr{\dd #1}{\dd #2}}
\def\dd{{\hbox{\rm d}}}
\def\ddovr#1#2{\ovr{\dd^2 #1}{\dd #2^2}}
\def\vec#1{\hbox{\bf #1}}
\def\Eq#1{Eq.~\eqref{#1}}
\def\biblist{\begingroup\advance\leftskip by 25pt 
	\parindent-25pt\frenchspacing\noindent}
\def\endbiblist{\par\endgroup}
\def\boldrule{\hrule height 1.2pt}
\def\noterule{\medskip\boldrule\medskip}	

\title{Natural Selection Maximizes Fisher Information}
\author{Steven A.\ Frank\footnote{Department of Ecology and Evolutionary Biology, University of California, Irvine, CA 92697--2525, USA, email: safrank@uci.edu}}
\begin{document}
\maketitle

\section*{Abstract}

In biology, information flows from the environment to the genome by the process of natural selection.   But it has not been clear precisely what sort of information metric properly describes natural selection.  Here, I show that Fisher information arises as the intrinsic metric of natural selection and evolutionary dynamics.  Maximizing the amount of Fisher information about the environment captured by the population leads to Fisher's fundamental theorem of natural selection, the most profound statement about how natural selection influences evolutionary dynamics.  I also show a relation between Fisher information and Shannon information (entropy) that may help to unify the correspondence between information and dynamics.  Finally, I discuss possible connections between the fundamental role of Fisher information in statistics, biology, and other fields of science.

\vfill
\noterule
\noindent{\bf Please cite as follows:} Frank, S. A. 2009. Natural selection maximizes Fisher information. Journal of Evolutionary Biology 22:231--244.
\noterule

\centerline{\bf The published, definitive version of this article is freely available at:} 
\medskip\centerline{\href{http://dx.doi.org/10.1111/j.1420-9101.2008.01647.x}{http://dx.doi.org/10.1111/j.1420-9101.2008.01647.x}}
\noterule

\vfill\eject

\begin{quotation}
\textit{Despite the pervading importance of selection in science and life, there has been no abstraction and generalization from genetical selection to obtain a general selection theory and general selection mathematics $\ldots$ Thus one might say that ``selection theory'' is a theory waiting to be born---much as communication theory was 50 years ago. Probably the main lack that has been holding back any development of a general selection theory is lack of a clear concept of the general nature or meaning of ``selection''$\ldots$\hfil\break\indent Probably the single most important prerequisite for Shannon's famous 1948 paper on ``A Mathematical Theory of Communication'' was the definition of ``information'' given by Hartley in 1928, for it was impossible to have a successful mathematical theory of communication without having a clear concept of the commodity ``information'' that a communication system deals with. (Price, 1995) }
\end{quotation}

\section*{Introduction}

Brillouin (1956) distinguished two types of information.  First, a natural phenomenon contains an intrinsic amount of information.  Second, observation transfers information about the phenomenon to the data.  Observations often do not completely capture all information in the phenomenon.  Wheeler (1992) and Frieden (2004) suggested that the particular form of dynamics in observed systems arises from the flow of information from phenomena to observation.  

These concepts of information flow seem similar to the process of evolutionary change by natural selection.  In biology, a population ``measures'' the intrinsic information in the environment by differential reproduction of individuals with varying phenotypes.  This fluctuation of phenotype frequencies transfers information to the population through changes in the frequencies of the hereditary particles.  However, the population does not fully capture all of the intrinsic information in the frequency fluctuations caused by differential reproduction, because only a fraction of phenotypic information flows to the next generation via changes in the frequencies of the hereditary particles.  

The analogy of natural selection as measurement and the analogy of heredity as the partial capture of the information in differential reproduction seem reasonable.  But how close can we come to a description of these processes in terms of a formal measure of information?  To study this question, I developed Frieden's (2004) conjecture that Fisher information is the metric by which to understand the relation between measurement and dynamics.  Fisher information provides the foundation for much of the classical theory of statistical inference (Fisher, 1922).  But the role of Fisher information in the conjectures of Brillouin, Wheeler, and Frieden remains an open problem.  

I show that maximizing the capture of Fisher information by the hereditary particles gives rise directly to Fisher's (1958) fundamental theorem of natural selection, the most profound statement of evolutionary dynamics.  I also extend Fisher information to account for observations correlated with the transfer of environmental information to the population, leading to a more general form that is equivalent to the Price equation for the evolutionary change in an arbitrarily defined character (Price, 1970, 1972a).  My analyses show that the Price equation and general aspects of evolutionary dynamics have a natural informational metric that derives from Fisher information. 

Although I show a formal match between Fisher information, the fundamental theorem of natural selection, and other key aspects of evolutionary analysis, I do not resolve several issues of interpretation.  In particular, it remains unclear whether the link between Fisher information and the evolutionary dynamics of natural selection is just an interesting analogy or represents a deeper insight into the structure of measurement, information, and dynamics.  My demonstration that the particular form taken by evolutionary dynamics arises naturally from Fisher information suggests something deeper, but the problem remains an open challenge.

\section*{Overview}

In the study of evolutionary dynamics, one must relate various quantities that change and the processes that cause change.  Consider, for example, the frequencies of hereditary units, such as genotypes, genes, vertically transmitted pathogens, and maternal effects.  The frequencies of such hereditary units may be influenced by selection, transmission, and mixing during recombination and sexual reproduction.  It is possible to express the relations between these quantities and processes in a wide variety of seemingly distinct ways, and with diverse notations.  However, a single formalism exists beneath all of those expressions.  That underlying formalism and several alternative expressions can be regarded as already reasonably well known.  

Why, then, should one look for new alternative expressions?  In my opinion, it is useful to understand more deeply the underlying formalism and to study the connections between evolutionary dynamics and concepts that have been developed in other fields of study.  The value of such understanding, based on analysis of alternative expressions, necessarily has a strongly subjective component.  There can never be a definitive argument against the claim that alternative expressions simply reformulate dynamics by notational change.  In defense of developing this work, one can reasonably say that natural selection is among the most profound processes in the natural world, and any further insight that can potentially be obtained about natural selection is certainly worth the effort.

To start, let us place natural selection in the context of evolutionary change.  The complexity of evolutionary dynamics arises in part from the three distinct levels of change.  First, characteristics, or phenotypes, determine how natural selection influences relative success, and the frequency distribution of character changes over time partly in response to differences in success associated with characters.   Second, individuals carry and transmit hereditary particles that influence their characters.  We usually associate hereditary particles with genes, but any transmitted entity could influence characters and could be tracked over time.  Third, each individual carries a particular array of hereditary particles, the array usually called the {\it genotype}.   

The relations between the genotype and the hereditary particles create one of the complexities of evolutionary dynamics.  In many organisms, individuals transmit a only subset of their hereditary particles, and combine their contribution of particles with another individual to make offspring with a new genotype.  Thus, differences in success do not directly change the frequency of genotypes. Rather, variation in genotypic success changes the frequency of the hereditary particles that are transmitted to make new combinations of genotypes in offspring.  

Further complexities arise because particles spontaneously change state (mutation), the effect of each particle varies with the combination of other particles with which it lives in genotypes, the environment changes through the changes in evolving organismal characters, and the environment changes for other reasons outside of the evolutionary system.  Because a vast number of outside processes potentially influence a particular evolutionary system, no complete description of evolutionary change can be developed from first principles.  Even within the confines of a closed system of evolutionary genetics, no description can both capture simple generalities and the complexities of each situation.  

In this paper, I focus on partial analyses of dynamics that highlight simple generalities common to all evolutionary systems.  I emphasize the role of natural selection in evolutionary change, because selection is perhaps the most important component of evolutionary change, and because one can draw some very clear conclusions about the role of selection in dynamics.  Other factors are often important, but do not lend themselves to simple and general insights about evolutionary dynamics. 

I start with the formal mathematical connections between Fisher information and the properties of natural selection, without much comment on deeper aspects of interpretation.  I then return to the problems of interpretation and alternative measures of information.

\section*{Fisher information}

Fisher information measures how much information observations provide about an unknown parameter of a probability distribution.  Suppose $p(y|\th)$ is a probability distribution function (pdf) of a random variable $y$ with parameter $\th$.  Define the log-likelihood of the parameter $\th$ given an observation of the random variable, $y$, as $L(\th|y) = \log[p(y|\th)]$, where I use $\log(\cdot)$ to denote the natural logarithm.  Then a simple univariate form of the Fisher information about the parameter $\th$ is
\begin{equation}\label{fisherInfo}
	F = \int_y \bigg(\dovr{L(\th|y)}{\th}\bigg)^2 p(y|\th) \dd y.
\end{equation}
This equation presents the standard form most commonly found in the literature.  I now present some variant forms to explain what Fisher information actually measures and to help in the study of evolutionary dynamics.

We can write an equivalent definition of Fisher information as 
\begin{equation}\label{fisherInfoSecond}
	F = -\int_y \bigg(\ddovr{L(\th|y)}{\th}\bigg) p(y|\th) \dd y,
\end{equation}
which measures the expected curvature, or acceleration, of the log-likelihood function with respect to the parameter $\th$ (see, for example, Arami \& Nagaoka, 2000).  A relatively flat log-likelihood surface means that an observation provides relatively little information about $\th$, whereas a strongly curved surface means that an observation provides relatively more information.

With regard to evolutionary dynamics, a more useful form emphasizes that Fisher information measures the distance between two probability distributions for a given change in the parameter $\th$.  To understand this distance interpretation, let us simplify the notation by defining $p_y=p(y|\th)$ and $p_y'=p(y|\th+\dd\th)$.  Let us also work with a discrete probability distribution, such that $y=1,\ldots,D$.  Then, an alternative form of \Eq{fisherInfo} can be written as 
\begin{equation}\notag
	F=\sum_y p_y\bigg(\dovr{\log(p_y)}{\th}\bigg)^2.
\end{equation}
Think of the probability distribution as a point in logarithmically scaled $D$-dimensional space, each dimension weighted by the square root of the frequency of that dimension in the initial distribution.  For the initial distribution at $\th$, the corresponding point in $D$ space is given by the vector $\vec{v}= \{\sqrt{p_y}\,\log(p_y)\}$ for $y=1,\ldots,D$; for the shifted distribution at $\th'=\th+\dd\th$, the corresponding point is given by $\vec{v}'= \{\sqrt{p_y}\,\log(p_y')\}$.  Then $F$ measures the square of the euclidean distance between these two vectors divided by $\dd\th^2$.  In other words, $F$ measures, on a logarithmic scale weighted by the initial frequencies, the squared euclidean distance between the distributions relative to the scale change defined by $\dd\th$.

A slight notational change emphasizes this interpretation of Fisher information as a distance between two distributions.  Write the slope of the log-likelihood function with respect to $\th$ as $\Ldot_y = \dd\log(p_y)/\dd\th = \dot{\log}(p_y)=\pdot_y/p_y$, where the overdot means differentiation with respect to $\th$.  Then
\begin{equation}\label{fisherInfoDot}
	F=\sum_y p_y\bigg(\ovr{\pdot_y}{p_y}\bigg)^2=\sum_y 
		p_y\Ldot_y^2=\sum_y\pdot_y\dot{\log}(p_y), 
\end{equation}
which emphasizes that Fisher information measures a squared distance between distributions when each dimension is scaled to give the fractional change in the distribution---logarithms are simply a scaling to measure fractional change.  This equation also gives us the first hint that Fisher information may be related to evolutionary dynamics: the $\pdot_y$ may be interpreted as the evolutionary change in frequency of the $y$th type in the population, where we may choose to label types by phenotype, genotype, or any other classification.

The rightmost form in \Eq{fisherInfoDot}, $\sum \pdot_y\dot{\log}(p_y)$, suggests that Fisher information is related to the Shannon information measure, or entropy, $-\sum p_y\log(p_y)$.  Later, I will show an equivalence relation between Fisher information and Shannon information with regard to the study of dynamics. 

\section*{Dynamics of type frequencies: selection}

I now relate Fisher information to one part of evolutionary dynamics: the change in the frequencies of types caused by variation in success.  I use the word {\it fitness\/} to denote a measure of success.  

In evolutionary dynamics, the fitness of a type defines the frequency of that type after evolutionary change.  Thus, we write $p_y' = p_y (w_y/\wbar)$, where $w_y$ is the fitness of the $y$th type, and $\wbar = \sum p_yw_y$ is the average fitness.  Recall that we may use $y$ to classify by any kind of type.

{\it It is very important to note the particular definition of fitness used in what follows.}  Here, $w_y$ is proportional to the fraction of the second population that derives from (maps to) type $y$ in the first population.  Thus, $p'_y$ does not mean the fraction of the population at $\th'$ of type $y$, but rather the fraction of the population at $\th'$ that derives from type $y$ at $\th$ (Frank, 1995, 1997; Price, 1995). 

The fitness measures, $w$, can be thought of in terms of the number of progeny derived from each type.  In particular, let the number of individuals of type $y$ at $\th$ be $N_y=Np_y$, where $N$ is the total size of the population.  Similarly, at $\th'$, let $N_y'=N'p_y'$.  Then $\wbar=N'/N$, and $w_y=N_y'/N_y$.

Fitness can alternatively be measured by the rate of change in numbers, sometimes called the Malthusian rate of increase, $m$.  To obtain the particular rates of change to analyze Fisher information, we see from \Eq{fisherInfoDot} that we need an expression for 
\begin{align}\label{aydef}
		\ovr{\pdot_y}{p_y} &= \logdot(p_y)\notag\\
					&= \logdot(N_y/N)\notag\\
					&= \logdot(N_y)-\logdot(N)\notag\\
					&= \dot{N}_y/N_y-\dot{N}/N\notag\\
					&= m_y - \mbar\notag\\
					&=a_y,
\end{align}
where $a_y$ is called the average excess in fitness (Fisher 1958; Crow \& Kimura, 1970).

Substituting into \Eq{fisherInfoDot} yields
\begin{equation}\label{popVar}
	P = \sum_y p_y\bigg(\ovr{\pdot_y}{p_y}\bigg)^2 = \sum_y p_ya_y^2,
\end{equation}
where $P$ denotes the total Fisher information in the population about $\th$ obtained from the frequency fluctuations $\pdot_y$.  Note that $P$ is also a normalized form of the total variance in fitness in the population, because $a_y^2$ is the squared fitness deviation of each type over the scale $\dd\th^2$.  The value of $P$ also denotes the square of a standardized euclidean distance between the initial population and the population after evolutionary change, as discussed in the section above on Fisher information. 

Frieden et al.\ (2001) noted that total evolutionary change, $P$, is a measure of Fisher information.  In particular, they wrote the total Fisher information as $F = \sum q_y(\qdot_y/q_y)^2$, where $\qdot=\dd q/\dd t$ is the time derivative of type frequencies.  Given that formulation, the Fisher information in the fluctuations of type frequencies provides information about time rather than about the environment.  However, Frieden et al.\ failed to advance beyond this stage of analysis because of their nonintuitive conclusion that information about time is the focal metric, and their lack of analysis with regard to the transmission of information via the hereditary particles.  To move ahead, I first show that time is not the correct metric about which information should be interpreted.  I then analyze the flow of information to the hereditary particles.

\section*{The scale of change}

Let us review Fisher information.  We begin with a probability distribution, $p$, that depends on some parameter, $\th$.  We may interpret the probability distribution as the frequency of various types in a population, for example, the frequency of genotypes or phenotypes.  The amount of Fisher information about $\th$ provided by an observation depends on how much the probability distribution, $p$, changes as $\th$ changes.  

The total Fisher information about $\th$ is equal to a measure of the squared distance, $\dd\log(p)^2$, that the probability distribution moves with respect to the scale of dynamics set by the underlying parameter, $\dd\th^2$.  From the definition given in \Eq{fisherInfoSecond}, Fisher information measures the observed acceleration of frequencies, and equates the observed acceleration to information about the unobserved force, $\th$.  

The typical view of evolutionary dynamics would be to consider $\th$ as a measure of time, and to follow the dynamics with respect to changes in time.  But a simple example suggests that using time alone as the scale of measure is not sufficient.  

Consider a standard dynamical equation for the change in frequency with respect to time, which takes the form
\begin{equation}\notag
	\dovr{p_y}{t} = s_y p_y.
\end{equation}
Here, $\dd t$ is the time scale over which change occurs, and $s_y$ is the rate at which change occurs for the $y$th type, subject to the constraint that the total change in frequencies is zero, $\sum\dd p_y=\sum s_y p_y\dd t = 0$.  The units on $\dd t$ are measured in time, $t$, and the units on the rate parameters, $s_y$, are measured in units $1/t$.  Note that each $s_y$ may depend on the frequencies of other types, thus this expression is a short-hand for what may be complex nonlinear dynamics and is only meant to emphasize the dimensions of change rather than the details of the dynamics.

Now make the substitution $s_y = s a_y$, yielding
\begin{equation}\notag
	\dovr{p_y}{t} = s p_y a_y.
\end{equation}
The dimensions on $s$ are $1/t$, thus $s\dd t=\dd\th$ is the dimensionless scale over which change is measured.  Because $p_y$ is a frequency, which is dimensionless, $a_y$ must also be dimensionless.  Thus, all dynamical expressions scaled over time have natural nondimensional analogs of the form
\begin{equation}\notag
	\pdot_y=\ovr{\dd p_y}{s \dd t} = \ovr{\dd p_y}{\dd\th} =p_ya_y.
\end{equation}

Similarly, we could measure change over space by $\dd L$, with dimensions $L$ for some length scale, and the rate of change with length by $\b$, with units $1/L$, so that $\b\dd L$ would be the dimensionless scale over which one measures dynamics.

These lines of evidence suggest that a proper interpretation of the dimensionless scale of change, $\dd\th$, would, for example, be $\dd\th = s\dd t$ or $\dd\th = \b\dd L$.  We may use other dimensionless scales, but in each case, the dimensionless quantity would be a change per unit scale multiplied by a length scale.  

Given the simple path from standard dynamics to the dimensionless interpretation of $\th$, why did it take me so long to arrive at this conclusion?  Because my point of departure was Fisher information, and I derived from Fisher information a view of dynamics as a distance between an initial population and a population changed over some arbitrary change in parameter, $\th$.  That path from Fisher information left the meaning of $\th$ unspecified, and so we had to connect a distance view of dynamics to a standard view based in time, and find $\th$ as a dimensionless scale of change.  

The path from Fisher information to dynamics raises another issue of interpretation.  In typical dynamical analyses, we start with given initial conditions and rates of change, and then calculate the changed population.  An analysis of evolutionary dynamics that begins with Fisher information follows a different path.  We start with observations of an initial population and a changed population, and use the observed distance between those populations to gain information about the unobserved environmental forces that determine the scale of change, $\th$.  Thus, in
\begin{equation}\notag
	\dd p_y = p_y' - p_y = p_ya_y\dd\th,
\end{equation}
the value of $a_y=\dd p_y/(p_y\dd\th)=\pdot_y/p_y$ measures the observed frequency perturbation over the unobserved scale of change $\th$. We use the observed set of frequency perturbations $\{a_y\}$ to obtain information about $\th$.  

To give an example, suppose we measure changes in frequency over time, so that $\dd\th = s\dd t$.  We can interpret $s$ as the environmental pressure for change per unit time, and $\dd t$ as the time change over which measurements occur.  The Fisher information that we obtain from frequency changes cannot separate between $s$ and $\dd t$; instead we only get information about the total dimensionless change, $\dd\th$.  

We can think of $\dd\th$ as the total pressure for change that the environment applies to the initial population, that is, a measure of the mismatch between the current population and the environment observed over the scale $\dd\th$.  Our measure of Fisher information based on observations about frequency fluctuations provides information about this mismatch.  

To summarize, if we start with observations of frequency fluctuations, then we can calculate information about the mismatch with the environment over the dimensionless scale $\dd\th$.  By contrast, if we start with initial frequencies and rules about change per some dimensional unit, we can calculate dynamics as altered frequencies per dimensional unit.  Thus, dynamics and information about environmental mismatch describe the same system, but represent different points of view with regard to what we observe and what we calculate from the given quantities.  Before we analyze these issues of interpretation more fully, it helps to consider from an informational perspective what else we can learn about natural selection.

\section*{Fisher information from correlated observations}

We often have measurements about the change in characters associated with fitness, such as weight, resistance to parasites, and so on.   In this section, I derive the Fisher information contained in observations that are correlated with fitness, where fitness is equivalent to frequency fluctuations.  In the next section, I use the result for correlated characters to place measures of Fisher information into the broader context of evolutionary dynamics given by the Price equation.

I begin with alternative forms of the total Fisher information in the population with respect to $\th$.  Start with the forms given by \Eq{popVar}  
\begin{equation}\label{sumSq}
	P = \sum_y p_y\bigg(\ovr{\pdot_y}{p_y}\bigg)^2 = \sum_y p_ya_y^2.
\end{equation}
Note that the form of each sum is
\begin{equation}\notag
	\sum_y p_y\d_y^2=\cov(\d,\d),
\end{equation}
because $\bar{\d}=0$ in each expression in \Eq{sumSq}.  Thus, we can, for example, write the total Fisher information from frequency fluctuations as
\begin{equation}\notag
	P=\cov(a,a).
\end{equation}
In general, we can write an equivalent expression for Fisher information based on any measurement $z$ correlated with $a$ as
\begin{equation}\notag
	R_{za}P=R_{za}\cov(a,a) =\cov(a,z),
\end{equation}
where $R_{za}$ is the regression of $z$ on $a$.  Note also that 
\begin{equation}\notag
	\cov(a,z) = \cov(m,z),
\end{equation}
because $a_y = m_y-\mbar$.  Thus, Fisher information can be expressed in terms of the covariance and the regression between fitness and an arbitrary character, $z$.  

It is also useful to note that the Fisher information that arises from natural selection of phenotypes, as expressed in the prior equation, can be expressed as a rate of change in the average value of a population.  First, note from earlier definitions of fitness that $w_y=1+m_y\dd\th$ and $\wbar=1+\mbar\dd\th$.  Thus, ignoring terms of order $\dd\th$, we have 
\begin{equation}\notag
	a_y=m_y-\mbar=(w_y/\wbar-1)/\dd\th,
\end{equation}
allowing us to expand the covariance between a character and fitness as
\begin{align}\label{zbarDiff}
		R_{za}P &= \cov(a,z)=\cov(m,z)\notag\\
					&= \sum_y p_y[(w_y/\wbar) -1]z_y/\dd\th\notag\\
					&= \Big(\sum_y p_y'z_y - \sum_y p_yz_y\Big)\Big/\dd\th\notag\\
					&= \sum_y \pdot_y z_y\notag\\
					&= \zdotbar_P,
\end{align}
in which the subscript $P$ denotes the focus on phenotypes and total population information.  Here, I do not account for any changes to the character during transmission or caused by changes in the environment.  I develop a full description of those additional processes of evolutionary change in the next section. 

\section*{The Price equation of evolutionary dynamics in terms of Fisher information}

Thus far, I have focused on the part of evolutionary change that follows from variation in fitness between types.  In particular, \Eq{sumSq} shows the equivalence between the variation in fitness, $\sum p_ya_y^2$, and frequency fluctuations given by the expression in terms of $\pdot_y$.  

In this section, I derive the Price equation, a general expression for total evolutionary change that accounts in an abstract way for all evolutionary forces, and for any arbitrary character (Price, 1970, 1972b; Frank, 1995, 1997).  I then use the Price equation to place natural selection and Fisher information into the wider context of evolutionary change.  The following sections use the Price equation to relate the change in the frequencies of the hereditary particles to the amount of Fisher information transmitted over the scale $\dd\th$.  That change in Fisher information through the hereditary particles leads us to the fundamental theorem of natural selection.

In what follows, let unprimed variables represent measurements at $\th$ and primed variables represent measurements at $\th' = \th+\dd\th$.  Overdots denote change with respect to $\th$ over the scale $\dd\th=\th'-\th$.  As before, a primed frequency, $p_y'=p_yw_y/\wbar$, represents the frequency of entities at $\th'$ derived from type $y$ at $\th$.  Similarly, for a measurement on types, $z_y$, the value $z_y'$ represents the average value of $z$ among those entities at $\th'$ derived from type $y$ at $\th$.  These special definitions for the mappings between two populations, first emphasized by Price (1995), give rise to a more general form of the Price equation first presented by Frank (1995, 1997).

With these definitions, the total change in the average value of some measurement, $z$, relative to the change between $\th'$ and $\th$, is
\begin{align}\label{PriceEq}
		\dot{\zbar} 	&= \Big(\sum p_y'z_y' - \sum p_yz_y\Big)\Big/\dd\th\notag\\
						&= \Big(\sum(p_yw_y/\wbar)(z_y+\dd z_y) - \sum p_yz_y\Big)
								\Big/\dd\th\notag\\
						&= \sum p_ya_yz_y + \sum p_y\zdot_y+ \sum p_y(w_y/\wbar-1)\dd z_y/\dd\th\notag\\
						&= \cov(m,z) + \E(\zdot)+\cov(m,\dd z)\notag\\
						&= \cov(m,z) + \E(\zdot),
\end{align}
where, in this continuous analysis, we can ignore $\cov(m,\dd z)$ as of magnitude $\dd z\rightarrow0$. \Eq{PriceEq} is a form of the Price equation.  We can express the second term alternately as 
\begin{equation}\label{EgivenP}
	\E(\zdot)=\sum p_y\zdot_y= \zdotbar_{E|P},
\end{equation}
where $\zdotbar_{E|P}$ is the rate of change in the character value that has nothing to do with changes in frequency, and thus has nothing do to with information about the environment transferred to the population via natural selection.  The subscript $E|P$ emphasizes that this environmental component arises in the context of analyzing phenotypes and total population information, $P$.

Putting the pieces together
\begin{align}\label{totalChange}
				\zdotbar &= \cov(m,z) + \E(\zdot) \notag\\
							&= \sum \pdot_y z_y + \sum p_y\zdot_y\notag\\
							&= R_{za}P + \zdotbar_{E|P}\notag\\
							& = \zdotbar_P + \zdotbar_{E|P}.
\end{align}
Note that the total change can be expressed in terms of pure Fisher information associated with the direct effects of frequency fluctuations, $\zdotbar_P$, and the environmental changes holding constant the consequences of frequency fluctuations caused by natural selection, $\zdotbar_{E|P}$.

Recall that Fisher information is a measure of distance between two probability distributions.  We see here that the distance metric in Fisher information can be translated into the rate of evolutionary change in mean character values caused by natural selection.  

\section*{Information in the hereditary particles}

The previous analysis did not explicitly consider the process by which characters transmit between populations.  Any changes in transmission are hidden within the environmental component of change, $\zdotbar_{E|P}$.  In biology, characters do not transmit as units, but rather as particles such as  genes and other transmissible units that contain information about the character.  Full analysis of evolutionary dynamics must consider the expression of information and change in terms of the transmissible, hereditary particles. 

In the previous section, I developed the equations for change in terms of an arbitrary character, $z$.  In this section, I focus on the character fitness, $m$, in order to connect the results to Fisher's famous fundamental theorem of natural selection, which is about the rate of change in fitness caused by natural selection (Fisher, 1958; Price, 1972a; Ewens, 1989, 1992; Frank \& Slatkin, 1992; Edwards, 2002).  I also introduce the hereditary particles into the analysis.  My analysis, focused on fitness, $m$, could be expanded to any character, $z$, correlated with fitness.

Let $y$ index subsets of the population, each subset containing a common set of hereditary particles.  Let the set defining $y$ be $\{x_{yj}\}$, where $j=1,\ldots,M$ labels the different kinds of hereditary particles that exist in the population.  Each type, $y$, contains $x_{yj}$ copies of particle $j$, and a total of $\sum_j x_{yj}=k$ particles.  

I now make a few additional assumptions that, while not necessary, lead to a convenient notation and to connections with classical population genetics.  Assume that each type has $L$ distinct slots (or loci), each slot containing $n$ particles, so that $k=nL$.  Further, assume that a particular particle can occur only at a particular slot.  Then the frequency of a particular particle is
\begin{equation}\notag
	r_j = \sum_y p_yx_{yj}/n.
\end{equation}
	
We measure the prediction about the character of interest provided by each particle by multiple regression (Fisher, 1958).  In this case, our focal character is fitness, $m$; as before, we use the average excess in fitness $a_y=m_y-\mbar$, and write
\begin{equation}\notag
	a_y = \sum_j \a_j x_{yj} + \e_y,
\end{equation}
where $\a_j$ is the partial regression coefficient of fitness on the predictor (particle) type $j$.  In genetics, $\a_j$ is called the {\it average effect} of the predictor (or allele) $j$ (Fisher, 1958; Crow \& Kimura, 1970).  If we sum both sides over the frequency of types, $p_y$, we have, by convention, $\sum p_y\e_y=0$, and from the definition of $r$, we obtain $\sum_j r_j\a_j = 0$.  

We can define $g_y = \sum_j \a_j x_{yj}$ as the fitness deviation predicted by the hereditary particles; the term $g_y$ is often referred to as the breeding value.  Thus, we can write
\begin{equation}\label{aydeftwo}
	a_y = g_y +\e_y, 
\end{equation}
which says that the observed fitness deviations and frequency fluctuations of types, $a_y = \pdot_y/p_y$, are equal to the deviations predicted by the hereditary particles carried by each type, $g_y$, and a deviation between the observed and predicted values, $\e_y$.  Because $g_y$ predicts the frequency fluctuations of types, we can write those predicted frequency fluctuations as $\gdot_y = (\g_y'-p_y)/\dd\th = g_yp_y$.  

How should the predictive value be chosen for the average effect of each hereditary particle, that is, how should values be assigned for the $\a_j$?  Following Fisher (1958), I choose the $\a_j$ to minimize the euclidean distance between observation and prediction, which means that the $\a_j$ are partial regression coefficients obtained by the theory of least squares.  Later, I will discuss the interpretation of why one would choose the value of the average effects of the hereditary particles with respect to this minimization criterion.  For now, I analyze the consequences, which relate to Fisher information and Fisher's fundamental theorem of natural selection.

The total squared distance between observation and prediction is the sum of the squared deviations for each type, each squared deviation weighted by the frequency of the particular type
\begin{equation}\label{LS}
	\sum_y p_y(a_y-g_y)^2 = \sum p_y\e_y^2.
\end{equation}
A geometric interpretation provides some insight into the consequences of minimizing this distance, and how these measures relate to Fisher information.  

\section*{A geometric interpretation of heredity and Fisher information}

Suppose there are $D$ different types, that is, $y=1,\ldots,D$.  Then the observed fitness deviations can be described as a point in $D$-dimensional space, $\bolda=\{\sqrt{p_y}a_y\}$, with a squared distance from the origin of 
\begin{equation}\label{totalFI}
	P=\sum p_ya_y^2=\sum p_y (\pdot_y/p_y)^2=\bolda\cdot\bolda,
\end{equation}
where, from \Eq{popVar}, we obtain the equivalence to the total Fisher information, $P$, and the sum of squared frequency deviations.  The dot product notation $\bolda\cdot\bolda$ expresses the sum of the product of each dimension between two vectors.  

Similarly, we can describe the predicted fitness deviations as a point $\boldg=\{\sqrt{p_y}g_y\}$, with a squared distance from the origin of 
\begin{equation}\label{geneticFI}
	G=\sum p_yg_y^2=\sum p_y (\gdot_y/p_y)^2=\boldg\cdot\boldg,
\end{equation} 
where, as defined above, $\gdot_y$ is the predicted frequency fluctuation of types.  Here, $G$ is a measure of Fisher information, because it matches the definition of Fisher information as a normalized distance between an initial probability distribution and an altered distribution, as in \Eq{fisherInfoDot}.

Returning to \Eq{LS}, we can express the minimization of the distance between observed frequency fluctuations, $a$, and predicted fluctuations, $g$, as a minimization of the distance between the points $\bolda$ and $\boldg$.  Because $\e_y=a_y-g_y$, we can express geometrically the deviations between observed and predicted fluctuations as $\bolde = \{\sqrt{p_y}(a_y-g_y)\}=\bolda-\boldg$.  Next, we can write in geometric notation the total squared distance between observation and prediction in \Eq{LS}, using the fact that $\bolda=\boldg+\bolde$, as 
\begin{align}\label{LSgeom}
			\bolde\cdot\bolde 	&= (\bolda-\boldg)\cdot(\bolda-\boldg)\notag\\
											&= \bolda\cdot\bolda - 2(\bolda\cdot\boldg)+\boldg\cdot\boldg\notag\\
											&= \bolda\cdot\bolda - 2(\boldg+\bolde)\cdot\boldg
																			+\boldg\cdot\boldg\notag\\
											&= \bolda\cdot\bolda - \boldg\cdot\boldg-2(\boldg\cdot\bolde).
\end{align}
The location of $\bolda$ is set by observation.  So, to minimize the squared distance between observation and prediction, we must choose the predicted values $\boldg$ to be as close to $\bolda$ as possible.  The standard theory of least squares, based on the principles of linear algebra, provides all of the details, which I will not derive here.  Rather, I will give the results, and briefly describe some intuitive ways in which those results can be understood.  The key is that the shortest distance between a point and a line is obtained by the perpendicular drawn from the point to the line.

Choosing $\boldg$ to be as close as possible to $\bolda$ means that the vector from the origin to $\boldg$ will be perpendicular to the vector between $\boldg$ and $\bolda$, denoted $\bolde$. Consequently $\boldg\cdot\bolde=0$.  The requirement that $\boldg$ be perpendicular to $\bolde$ can be understood as follows.  Since $\bolda$ is fixed in location, the distance between $\bolda$ and $\boldg$ depends on the length of $\boldg$ and the angle between $\bolda$ and $\boldg$ drawn as vectors from the origin.  

Recall the definition of $g$ given by $g_y = \sum_j\a_jx_{yj}$. The $x$ values are fixed, but we are free to choose the $\a_j$ in order to make $\boldg$ as close as possible to $\bolda$, subject to the one constraint that $\sum_y p_yg_y = \sum_j r_j\a_j = 0$.  This constraint is still satisfied if we multiply all $\a_j$ values by a constant.  Thus, we can freely choose the length of the vector $\boldg$, assuming that we have at least one degree of freedom in setting $\boldg$ after accounting for the single constraint on the $\a_j$.  If we can set the length of $\boldg$, then no matter what constraints there are on reducing the angle between $\bolda$ and $\boldg$, the shortest distance between $\bolda$ and $\boldg$, given by the vector $\bolde=\bolda-\boldg$, occurs when the vector $\bolde$ is perpendicular to $\boldg$.  

Correlations in the $x$ values can lower the number of degrees of freedom we have for reducing the angle between observation, $\bolda$, and prediction, $\boldg$, causing an increase in the minimum distance between observation and prediction.  However, such correlations between the $x$ values do not prevent adjusting the length of $\boldg$ as long as there is one degree of freedom available.  Thus, correlated predictors (alleles) do not alter the conclusion that $\boldg\cdot\bolde=0$.  In terms of classical population genetics, nonrandom mating and linkage between different alleles may cause correlations between the alleles.  The point here is that such processes do not alter the conclusion that minimizing the distance between observation, $\bolda$, and prediction, $\boldg$, leads to $\boldg\cdot\bolde=0$.  Using this fact in \Eq{LSgeom},
\begin{equation}\notag
	\bolde\cdot\bolde=\bolda\cdot\bolda - \boldg\cdot\boldg.
\end{equation}
Recall from \Eq{totalFI} that $P = \bolda\cdot\bolda$, the intrinsic or total Fisher information about the environment measured by the population in regard to the observed fluctuations of phenotypic fitness.  Similarly, from \Eq{geneticFI}, $G=\boldg\cdot\boldg$, the Fisher information about fitness captured by fluctuations in the hereditary particles, when the average effect of each particle is obtained by minimizing the distance between the observed and predicted fitnesses.  If we define $E=\bolde\cdot\bolde$ as the difference between $P$ and $G$, then we can express the distance relations of \Eq{LSgeom} in terms of Fisher information as
\begin{equation}\notag
	P-G=E,
\end{equation}
where $G$ is the portion of the total Fisher information, $P$, captured by the hereditary particles, and $E$ is the portion of the Fisher information lost by the population.  In terms of traditional genetics, $P$ is the total variance in the population, $G$ is the genetic variance, and $E$ is the environmental variance, with $G/P$ a measure of heritability (Crow \& Kimura, 1970).

The new result here is that these traditional measures of variance are measures of Fisher information.  I suggest that the role of variances in the fundamental equations of biology can be understood as measures of Fisher information.

\section*{The total change in fitness}

I prepare my exposition of the fundamental theorem of natural selection by first expressing the components of the total change in fitness. I use \Eq{totalChange} as a starting point, then transform the components into terms that can be ascribed to the transmissible particles.  For example, I measure the direct effect of natural selection as the change in the character that arises from the changed frequency of the transmissible particles caused by natural selection.  In this case, I analyze fitness itself as the character of interest.

We can express the fitness of the $y$th type in terms of the predicted fitness deviation, $g_y$, and the residual, $\e_y$, by combining \Eq{aydef} and \Eq{aydeftwo}, yielding
\begin{equation}\notag
	m_y = \mbar+ g_y+\e_y.
\end{equation}
The predicted value of the fitness for the $y$th type is obtained by dropping the residual term and using only the predicted deviations, $g_y$, yielding the predicted fitness
\begin{equation}\notag
	v_y = \mbar+ g_y.
\end{equation}
With these definitions, we can analyze the total change in fitness over the scale $\dd\th$ by focusing on the predicted fitnesses given by $v$, because
\begin{equation}\notag
	\mdotbar = (\mbar'-\mbar)/\dd\th=\Big(\sum p_y'v_y'-\sum p_yv_y\Big)
			\Big/\dd\th=\vdotbar,
\end{equation}
the equality derived by the fact that $\sum p_y'g_y' = \sum p_yg_y=0$.  From \Eq{totalChange}, using $v$ in place of $z$, we can write
\begin{equation}\notag
	\mdotbar = \vdotbar=\sum \pdot_y v_y + \sum p_y\vdot_y.
\end{equation}
Next, expand the first term on the right, using the fact from \Eq{zbarDiff} that, for $z\equiv g$, we can write $\sum \pdot_y g_y = \cov(m,g)$, thus
\begin{align}\notag
		\sum\pdot_y v_y 	&= \sum\pdot_y g_y\notag\\
									&= \cov(m,g)\notag\\
									&= \cov(g+\e,g)\notag\\
									&= \cov(g,g)\notag\\
									&= G\notag\\
									&=\mdotbar_G,\notag
\end{align}
where $\mdotbar_G$ arises by analogy with \Eq{zbarDiff}; here the subscript $G$ denotes that this term quantifies the rate of change in fitness explained directly by the predicted fitness deviations, $g$, based on the hereditary particles.  

The second term, from \Eq{EgivenP}, is
\begin{equation}\notag
	\sum p_y\vdot_y = \mdotbar_{E|G},
\end{equation}
where the subscript $E|G$ denotes environmental changes that alter the transmission of character value independently of changes in frequencies caused by selection.  Here, the context is $G$, because the value measured by $v$ arises as a prediction based on the hereditary particles, rather than the actual value.

Putting the pieces together, we can express the rate of total change in fitness as
\begin{equation}\label{totalFitChange}
	\mdotbar = \mdotbar_G + \mdotbar_{E|G}. 
\end{equation}
	
\section*{The fundamental theorem of natural selection}

Fisher (1958) partitioned the total change in fitness into two components.  First, he ascribed to natural selection the part of total change caused by differential success of types and consequent change in the frequency of the hereditary particles (Price, 1972b; Ewens, 1989).  Fisher calculated this natural selection component by ignoring any change in the average effects of the hereditary particles, because he assumed that such changes were not caused directly by natural selection.  Second, following Fisher, I have called any component that depends on changes in the average effects of the hereditary particles a component of environmental change (Frank \& Slatkin, 1992). 

Holding the environment constant and thereby ignoring any changes in the average effects of the hereditary particles, $\mdotbar_{E|G}=0$, we have, from the previous section, Fisher's fundamental theorem of natural selection as the fisherian partial change in fitness ascribed to natural selection
\begin{equation}\label{FTNS}
	\mdotbar_f = \mdotbar_G = G,
\end{equation}
where the subscript $f$ denotes the fisherian partial change.  Fisher expressed $G$ as the genetic variance in fitness, which we can see from the definition of $G$ in \Eq{geneticFI}, because the average value of $g$ is zero.  

I showed that $G$ is also the Fisher information about fitness captured by the hereditary particles.  \textit{Thus, the fundamental theorem can be expressed as: the rate of change in fitness caused by natural selection is equal to the Fisher information about the environment captured by the hereditary particles.}

\section*{The fundamental theorem and frequency fluctuations of hereditary particles}

The fundamental theorem expresses that part of total change in fitness caused by changes in the frequencies of the hereditary particles.  However, the statement of the theorem in \Eq{FTNS} does not show explicitly how $G$ is related to the changes in the hereditary particles.  In this section, I make explicit the relation between the Fisher information captured by the hereditary particles and the changes in the frequencies of the hereditary particles.

Begin, from \Eq{geneticFI}, by noting that the Fisher information captured by the population is written as
\begin{equation}\notag
	G=\sum p_yg_y^2=\boldg\cdot\boldg,
\end{equation} 
where $\boldg=\{\sqrt{p_y}g_y\}$ is a point in $D$-dimensional space over $y=1,\ldots,D$.  Rewrite this expression for $G$ as
\begin{equation}\notag
	\sum p_yg_y^2 = \cov(g,g)=\cov(a,g) = \sum p_y a_y g_y = \sum \pdot_y g_y,
\end{equation}
because $a=g+\e$, and $\cov(\e,g)=0$.  Now,
\begin{align}\notag
		\sum_y \pdot_y g_y &= \sum_y \pdot_y\sum_j x_{yj}\a_j\notag\\
										&= \sum_j\Big(\sum_y\pdot_y x_{yj}\Big)\a_j.\notag
\end{align}
Next, expand the inner summation
\begin{align}\notag
		\sum_y\pdot_y x_{yj} &= \Big(\sum_y p_y'x_{yj}- \sum_y p_yx_{yj}\Big)
											\Big/\dd\th\notag\\
										&= n(r_j' - r_j)/\dd\th\notag\\
										&=n\rdot_j,\notag
\end{align}
where I assume that any change in $x_{yj}$ in the descendant is ascribed to a change in the environment, because changes in the state of hereditary particles are not direct consequences of natural selection. Put another way, $x'_{yj}=x_{yj}$, where $x'_{yj}$ denotes the hereditary particle derived from $x_{yj}$.  If the actual state of the hereditary particle at $\th'$ differs from $x_{yj}$, then we account for this through the change in the average effect (see \Eq{particleB} below).

Putting the pieces together
\begin{equation}\label{particleFI}
	G=n\sum_j \rdot_j\a_j=\boldg\cdot\boldg, 
\end{equation}
which means that we can alternatively express the location of the predicted fitness fluctuations as $\boldg = \{\sqrt{n\rdot_j\a_j}\}$ in the $M$-dimensional space over $j=1,\ldots,M$.  If the particles, $x$, are correlated within individuals, then the $\rdot_j$ may be correlated such that $\boldg$ is confined to a subspace of lower dimension than $M$.  Such correlation may arise in biology from nonrandom mating or linkage.  Correlation would not affect any conclusion, but may force the predicted fitness fluctuations, $\boldg$, to be farther from the observed fitness fluctuations, $\bolda$.  

The previous section showed that, by the fundamental theorem, the rate of change in fitness captured by the hereditary particles is the Fisher information captured by the hereditary particles, $G$.  We see in \Eq{particleFI} a more direct expression of $G$ in terms of the fluctuations in the frequencies of the hereditary particles.

An alternative form for the average effects sets the partial change of the fundamental theorem in the context of total evolutionary change.  Express the average effect of a hereditary particle as $b_j=\mbar(\ovr{1}{n}+\a_j)$, and the effect in the changed population as $b_j'=\mbar'(\ovr{1}{n}+\a_j')$; thus, $\mdotbar=n\dot{\bar{b}}$.  We can, by analogy with \Eq{totalChange}, express the total change in fitness as
\begin{align}\label{particleB}
		\mdotbar &= n\Big(\sum \rdot_j b_j + \sum r_j \bdot_j\Big)\notag\\
						&= G + n\sum r_j \bdot_j. 
\end{align}
The first part, $G$, is the partial change ascribed to natural selection---this is the part that comprises the fundamental theorem.  The second part provides an explicit description for the remaining part of total change.

If we assume that the hereditary particles have constant effects on fitness, we obtain $\bdot_j=0$.  By contrast, constant effects lead to $\adot_j=-\mdotbar/n$, because the constraint that $\sum r_j\a_j=0$ forces the average effects to adjust for changes in the frequencies of the particles and consequent change in mean fitness.  Thus, to study the total change in fitness, $b_j$ is a more natural metric.  For example, changes in the $b_j$ may reflect true changes in effects in response to changes in particle frequencies, rather than adjustments for changes in the mean.

The fisherian partial change in fitness arises through the frequency changes of particles when holding constant the average effects.  Thus, the $\bdot_j$ describe all aspects of change other than the fisherian partial change in fitness.  When deciding how to choose the set of hereditary particles, a natural approach would be to increase $G$ and reduce the second term on the right of \Eq{particleB}; in other words, one might choose the most stable set of hereditary particles that explain the largest fraction of the total change through $G$.   This criterion provides guidance for whether to consider nongenetic factors as hereditary particles.

\section*{Scope of the fundamental theorem}

In traditional population genetics, one usually specifies at the outset particular assumptions about how individuals mate, how genes mix, and how genes are linked to each other.  The scope of derivations from such assumptions remains limited to the particular systems of mating and mixing specified.  By contrast, I did not make any specific assumptions about mating and mixing.  Rather, my derivation transcends traditional genetics and applies to any dynamical system with a clear notion of success based on selection.  My hereditary particles are just any predictors that can be identified and associated with success in selection---that is, with fitness.  

I divided the original population into subsets indexed by $y$.  We may consider the index $y$ to be different individuals, different genotypes, or any other partition---it does not matter.  I defined the frequency of type $y$ in the original population as $p_y$.  The expression $p_y'=(w_y/\wbar)p_y$ defines fitness, where $p_y'$ is the fraction of the descendant population derived from members of the class given by index $y$ in the original population.  Thus, fitness is simply a descriptive mapping between two populations.  If two or more sexes mix their hereditary particles to make each descendant, then we assign to each parental contributor a fraction of the descendant.  

We may, for example, consider subsets of the set $y=1,\ldots,D$ to be different sexes or different classes of individuals, each sex or class with potentially a different net contribution to the subsequent population.  In population genetics, we call the net contribution of each class the class reproductive value.  However, all of those details automatically enter into the values of fitness, $w_y$, because I have defined $w_y$ to be a mapping that measures fully the contribution of $y$ to the descendant population.  Complex mappings between populations may require a finely divided classification by $y$. But the system works for any pair of populations for which one can draw a map that associates components of each descendant to an ancestral entity.  

Another commonly discussed issue concerns correlation, or linkage, between hereditary particles.  My geometric expressions for the population distribution of fitnesses and the predicted fitnesses showed that correlation between hereditary particles may constrain the location of the vector that characterizes the predicted fitnesses---or, we may say equivalently, the predicted frequency fluctuations.  But such constraints do not change the fundamental properties by which distance is minimized in setting the predictor vector and determining the orthogonal (uncorrelated) directions of the prediction and mismatch vectors, $\boldg$ and $\bolde$.

In short, my derivation applies to any selective system for which a proper mapping between populations can be defined to express the fitness relations, $p_y'=(w_y/\wbar)p_y$.  If one can express the fractions of the descendant population that derive from the ancestral population, then the fundamental theorem follows.  This expression, in its abstract form, is a general description of selection for any system.  

What about the hereditary particles?  All that we need is to express a prediction about fitness, $w_y$, in terms of some predictors associated with $y$.  The theorem works with any set of predictors.  But, with such abstract generality, there is no single realization of the fundamental theorem, because there is no single set of predictors or hereditary particles that exist.  One can use alleles and follow allele frequency changes, as Fisher did.  But one could also include inherited pathogens, maternal effects, cultural predictors however defined, or the interaction effects predicted by combinations of predictors.  

The natural set of hereditary particles should probably be those predictors that remain most stable during transmission; otherwise, the changes caused by selection disappear in the descendant population, because large changes in average effects cause much of total change to be ascribed to the environment according to \Eq{particleB}.  One might define an optimality criterion to delimit the set of predictors and hereditary particles with regard to stability of the particles and the amount of Fisher information captured by the particles.

From a different perspective, one may analyze how the effectiveness of natural selection rises with an increase in the stability of the hereditary particles.  It may be that natural selection itself favors an increase in the stability of the hereditary components, thereby separating the rate of change between selective and environmental components of evolution.  Such time scale separation forms the basis for the subject of niche construction (Odling-Smee et al., 2003; Krakauer et al., 2008).

\section*{Fisher information, measurement, and dynamics}

Fisher information fits elegantly into a framework of natural selection and evolutionary dynamics.  But is the fit of Fisher information with evolutionary dynamics truly meaningful, or is the fit simply an outcome of altered notation?

I answer in two ways.  First, Fisher information does express evolutionary dynamics by a simple alteration of notation.  This must be so, because I showed that Fisher information provides a measure of change, and change arises from dynamics.  

Second, although Fisher information describes dynamics, it also represents a different perspective.  Typically, in the study of dynamics, one begins with initial frequencies over states, and rules for change.  From the initial conditions and rules for change, one derives the changed frequencies.  By contrast, Fisher information takes the initial and changed populations as given, and asks how much information can we obtain from the observed changes about the selective processes that determine those changes.  

Put another way, we can think of dynamics as depending on three variables: initial state, changed state, and rules that determine change.  Given any two, we gain information about the third.  

Natural selection is a rule that governs change.  We generally cannot observe or directly measure natural selection.   We can only infer natural selection from observed frequency fluctuations.  For this reason, Fisher information is a very clear and important perspective on natural selection: Fisher information measures how much we can learn about the unobserved selective processes of nature from the observed frequency fluctuations.  

By contrast, traditional dynamical theory starts with an initial state and a hypothesized rule for change imposed by selection, and then makes a prediction about the changed state that can be compared with observation.  

There is no a priori reason to conclude that the traditional dynamical theory is better or worse than the Fisher information perspective.  Each emphasizes a different view of dynamics.  I prefer the Fisher information perspective, because we can say exactly how much Fisher information about natural selection we can obtain from observed frequency fluctuations and a particular set of hereditary particles.  That seems like a statement of greatest generality that can be used to understand natural selection.  

By contrast, from the traditional dynamical view, all we can say is that some part of the total change in fitness is caused by natural selection and there is some remainder term.  That may be useful in some situations, but it does not seem to be very general or powerful with regard to studying natural selection.  The limitation of the traditional dynamical view probably explains why Fisher's fundamental theorem has been almost universally misinterpreted.

\section*{Shannon information compared with Fisher information}

Fisher never related his fundamental theorem of natural selection to Fisher information.  When he presented his theorem, he did draw an analogy between the fundamental theorem and the second law of thermodynamics.  Following Fisher, some have tried to relate natural selection to thermodynamics and measures of information, but never with much success.  I think the problem arose because thermodynamics suggested entropy and measures of information from the Shannon index family rather than Fisher information.  I do not know of anyone who has clearly related Shannon information and entropy to evolutionary dynamics and Fisher's fundamental theorem.

We have seen that Fisher information provides a natural metric for evolutionary dynamics; the question here concerns how Shannon information and entropy relate to Fisher information and evolutionary dynamics.  In this section, I derive a simple relation between Shannon information and Fisher information.  

Entropy in physics is defined as $S=-\sum p_y\log(p_y)$; the Shannon index of information has the same definition but is usually denoted by $H$.  A vast literature debates whether $S$ and $H$ are conceptually equivalent or, alternatively, whether the corresponding forms of $S$ and $H$ are merely coincidental (Ben-Naim 2008).  The relations between information and thermodynamics do not concern me here; I focus on the relations between Shannon information and Fisher information in the context of understanding dynamics.

The component of evolutionary dynamics ascribed to natural selection, $G$, arises by maximization of Fisher information.  Fisher information measures the observed acceleration in the frequencies of a probability distribution with respect to some unobserved parameter (force); from the observed acceleration, one obtains information about the unobserved force.  These relations can be seen in the definition of the Fisher information about a parameter $\th$, given earlier as
\begin{equation}\label{fisherInfoAcc}
	F = -\sum p(y|\th)\bigg(\ovr{\dd^2 L(\th|y)}{\dd\th^2}\bigg), 
\end{equation}
where $L = \log[p(\th|y)]$ is the log-likelihood of $\th$ given an observed value of $y$.  

I now propose an interpretation by which Fisher information is equivalent to the acceleration of Shannon information.  I show that, if we differentiate Shannon information twice with respect to a parameter, we obtain Fisher information.  However, this correspondence only arises if we assume a particular interpretation of what it means to measure the acceleration of Shannon information with respect to a parameter.  

To begin, let us consider Shannon information from the same perspective as Fisher information: how much do we learn about a parameter, $\th$, from an observation of the random variable, $y$?   We can rearrange Shannon information as
\begin{align}\notag
			H 	&= -\sum p(y|\th)\log[p(\th|y)]\notag\\
				&= -\sum p(y|\th)L(\th|y).\notag
\end{align}
To obtain the acceleration of Shannon information, we differentiate $H$ twice with respect to $\th$.  To differentiate, we may consider two choices.  

First, we may think of both $p(y|\th)$ and $L(\th|y)$ as functions of $\th$, and apply the chain rule.  This approach makes sense if we wish to study the total change in the information or entropy measure between two different distributions.  We may, for example, wish to find a distribution that maximizes the total entropy. To find such a distribution, we would need to study the total entropy change between different distributions.

Second, we may regard $p(y|\th)$ differently from $L(\th|y)$ with respect to the parameter $\th$.  In particular, $-L(\th|y)$ may be thought of as the information that we obtain about the variable $\th$ given an observation $y$.  In the theory of Shannon information, $-L(\th|y)$ is referred to as ``self-information'': a measure of the information obtained from the observation of a particular value of a random variable.  Tribus (1961) called $-L(\th|y)$ the ``surprisal'': a measure of the surprise in observing a particular outcome, $y$.  

By these interpretations of $-L$ as the measure of information in an observation, we obtain the expected surprise, or Shannon information, $H$, by averaging $-L$ over the distribution $p(y|\th)$.  Thus, if we want a measure of the expected second derivative, or acceleration, of the information in an observation, it makes sense to hold $p(y|\th)$ constant, and differentiate only the information, $-L$.  

If we differentiate $-L$ twice with respect to $\th$, and then average over the distribution $p(y|\th)$, we obtain the acceleration of Shannon information, $\ddot{H}$, as the expected acceleration of $-L$.  In this case, the acceleration of Shannon information equals Fisher information, as in \Eq{fisherInfoAcc}. 

From \Eq{fisherInfoSecond} and \Eq{fisherInfoDot}, the expected acceleration of the likelihood is
\begin{equation}\notag
	\ddot{H}=F =  \sum p_y\bigg(\ovr{\pdot_y}{p_y}\bigg)^2.
\end{equation}
We may read this as: the acceleration of information with respect to $\th$ is the squared distance between probability distributions over the scale $\th$, taken from the perspective of the distribution at the point $\th$.  Taking the perspective at $\th$ means that we weight the distances at each value $y$ by the probability $p(y|\th)$.  

This measure of acceleration provides a natural correspondence between dynamics and information.  From observed dynamics, $\pdot_y$, we learn information about an unobserved parameter, $\th$.  Put another way, from observed acceleration, $(\pdot_y/p_y)^2$, we learn about an unobserved force embedded in the scale $\th$.  The correspondence between acceleration and force is the fundamental principle behind all dynamics. 

Frieden (2004) has shown that many aspects of dynamics in physics can be derived from a principle that in effect maximizes the Fisher information in the frequency fluctuations that characterize dynamics.  With regard to Shannon information, and the equivalent measure of entropy, Martyushev \& Seleznev (2005) review many lines of evidence that dynamical trajectories follow paths that maximize the gain of entropy or Shannon information.  

In both Frieden (2004) and Martyushev \& Seleznev (2005), the fundamental correspondence between information (entropy) and dynamics arises by variational principles, in which dynamical paths and, equivalently, information measures, are extremized subject to external conditions imposed.  I have shown that, with regard to force, acceleration, and dynamics, Fisher information and Shannon information are equivalent.

So, in the end, we return to the question I posed in the introduction.  Is the correspondence between evolutionary dynamics and information fundamental and useful to our way of thinking?  I believe that dynamics and information are two alternative perspectives of the same phenomenon: the dynamical view begins with an observed or supposed force and deduces acceleration; the informational view begins with observed or supposed acceleration and induces force.

\section*{Maximization of Fisher information in biology and other sciences}

Frieden (2004) has argued that Fisher information is the natural metric for all of the sciences.  Frieden's own work has concentrated primarily on physics.  I know little of physics, and so I cannot comment extensively on Frieden's work and its relation to my discussion of Fisher information in evolutionary dynamics.  However, I did get the idea of applying Fisher information to natural selection from reading Frieden.  The most important point of connection between Frieden's physics and my analysis arises from the role played by the maximization of Fisher information.

In Frieden's framework, the physical constraints that define the dynamics of a particular natural phenomenon contain an intrinsic amount of information, $J$.  Observation of the dynamics, measured in terms of frequency fluctuations, transfers information about the phenomenon to the data, yielding a level of information in the data about the phenomenon, $I$.  Observations may not completely capture all information in the phenomenon, thus $I\le J$; we can write $J-I=-K$, where $-K\ge0$ is the information lost.  If one quantifies the informational measures in terms of Fisher information, then Frieden shows that physical phenomena typically minimize $-K$.  Minimization of $-K$ means that measurement transfers the maximum amount of Fisher information from $J$ to $I$, that is, from the phenomenon to the data.  

In most physical problems, the bound information $J$ is not observed directly.  Instead, $J$ acts as the unobserved source for the information $I$ received in the data.  The value of $J$ must be inferred, such that the information in observed frequency fluctuations, $I$, plus minimization of $-K$ derives the correct description of dynamics.  

In the biological problem that I analyzed, the bound information about the environment, $P\equiv J$, arises from the total frequency fluctuations observed in the population; the captured information, $G\equiv I$, arises from weighting the frequency fluctuations of the hereditary particles so as to maximize the prediction of the total population fluctuations; and the information lost is $E\equiv-K$.  Maximizing the prediction inherent in $G$ is equivalent to minimizing $P-G=E$, or, equivalently, to maximizing the Fisher information $G$ subject to the fixed value of $P$ set by observation.  Thus, my analysis of natural selection has a similar structure as Frieden's method, but differs with regard to how one interprets the bound information and the information in the data.  

Leaving aside physics, how should we interpret the maximization of Fisher information in our understanding of natural selection?  This question leads us back to the two alternative frames of reference with respect to dynamics, each frame with distinct and complementary lessons.

The traditional view of dynamics is: an initial probability distribution over states plus rules of change lead to a prediction of an altered distribution over states.  One analyzes the quality of the hypothesized rules of change by the distance between the predicted and actual distribution over observed states after dynamical change.  This view is inherently deductive:  one arrives at the rules of change by deducing them from extrinsic principles or hypotheses.  One then tests the deductive predictions against the observed distribution after dynamical change.

The Fisher information view of dynamics is:  the distance between the initial and subsequent distributions over states provides information about the unobserved rules of change.  One measures the quality of the information about the unobserved rules by Fisher information.  This view is inherently inductive: one arrives at estimates of the rules of change by iterative accumulation of information measured at particular points, weighed against decay of information as the rules change with respect to the points of measurement.  In biology, the hereditary particles, or predictors, are the stores of information, and represent inductively achieved hypotheses that are tested in each round of measurement.

Fisher information seems the perfect framework for analyzing the role of natural selection in evolutionary dynamics.  Natural selection must accumulate information about the environment inductively, acquiring information by changes in the frequencies of the hereditary particles.  The interesting problem concerns how much of the total information about the environment, $P$, transfers to the population through the information gain by the hereditary particles, $G$.  I showed that Fisher's fundamental theorem of natural selection follows from the assumption that $G$ contains the maximum amount of Fisher information about the environment that can be captured in the frequency fluctuations of the hereditary particles.

Why should $G$ be maximized?  Because the population will dynamically move toward its maximum fitness at a rate predicted exactly by the maximization of $G$, given a set of hereditary particles that remain stable in their average effect and do not change in their predictions (or causes) of fitness (Robertson, 1966; Crow \& Nagylaki, 1976; Ewens, 1992).  Thus, in the absence of any evolutionary force other than natural selection, we see the direct effect of natural selection and its action in maximizing the accumulation of Fisher information with respect to the hereditary particles.

This line of thought does not in any way require that the hereditary particles actually remain stable.  In any realistic situation, the effects of the hereditary particles will change, for example, if the effects depend on the frequencies of the particles.  The argument does show how to isolate the direct effect of natural selection.  That direct effect always moves the population in the direction of increasing fitness at a rate that arises from the maximization of Fisher information captured by the hereditary particles.

\section*{Acknowledgments}

I am grateful to Warren Ewens for our many lively discussions over the past 20 years about R.~A. Fisher and George Price.  Roy Frieden generously answered my many questions about Fisher information and dynamics.  National Institute of General Medical Sciences MIDAS Program grant U01-GM-76499 supports my research.

\vfill\eject

\section*{References}

{\biblist

Amari, S. \& Nagaoka, H. 2000. Methods of Information Geometry. American Mathematical Society, Providence, RI.

Ben-Naim, A. 2008. A Farewell to Entropy: Statistical Thermodynamics Based on Information. World Scientific Publishing, Hackensack, New Jersey.

Brillouin, L. 1956. Science and Information Theory. Academic Press, New York.

Crow, J. F. \& Kimura, M. 1970. An Introduction to Population Genetics Theory. Harper and Row, New York.

Crow, J. F. \& Nagylaki, T. 1976. Rate of change of a character correlated with fitness. American Naturalist 110: 207-213.

Edwards, A. W. F. 2002. The fundamental theorem of natural selection. Theoretical Population Biology 61: 335-337.

Ewens, W. J. 1989. An interpretation and proof of the fundamental theorem of natural selection. Theoretical Population Biology 36: 167-180.

Ewens, W. J. 1992. An optimizing principle of natural selection in evolutionary population genetics. Theoretical Population Biology 42: 333-346.

Fisher, R. A. 1922. On the mathematical foundations of theoretical statistics. Philosophical Transactions of the Royal Society, A 222: 309-368.

Fisher, R. A. 1958. The Genetical Theory of Natural Selection (2nd ed.). Dover Publications, New York.

Frank, S. A. 1995. George Price's contributions to evolutionary genetics. Journal of Theoretical Biology 175: 373-388.

Frank, S. A. 1997. The Price equation, Fisher's fundamental theorem, kin selection, and causal analysis. Evolution 51: 1712-1729.

Frank, S. A. \& Slatkin, M. 1992. Fisher's fundamental theorem of natural selection. Trends in Ecology and Evolution 7: 92-95.

Frieden, B. R. 2004. Science from Fisher Information: A Unification. Cambridge University Press, Cambridge, UK.

Frieden, B. R., Plastino, A. \& Shoffer, B. H. 2001. Population genetics from an information perspective. Journal of Theoretical Biology 208: 49-64.

Krakauer, D. C., Page, K. \& Erwin, D. 2008. Diversity, dilemmas and monopolies of niche construction. American Naturalist (in press).

Martyushev, L. M. \& Seleznev, V. D. 2006. Maximum entropy production principle in physics, chemistry and biology. Physics Reports 426: 1-45.

Odling-Smee, J., Laland, K. \& Feldman, M. 2003. Niche Construction: The Neglected Process in Evolution. Princeton University Press, Princeton.

Price, G. R. 1970. Selection and covariance. Nature 227: 520-521.

Price, G. R. 1972a. Extension of covariance selection mathematics. Annals of Human Genetics 35: 485-490.

Price, G. R. 1972b. Fisher's `fundamental theorem' made clear. Annals of Human Genetics 36: 129-140.

Price, G. R. 1995. The nature of selection. (Written circa 1971, published posthumously). Journal of Theoretical Biology 175: 389-396.

Robertson, A. 1966. A mathematical model of the culling process in dairy cattle. Animal Production 8, 93-108.

Tribus, M. 1961. Thermostatics and Thermodynamics: An Introduction to Energy, Information and States of Matter, with Engineering Applications. Van Nostrand, New York.

Wheeler, J. A. 1992. Recent thinking about the nature of the physical world: It from bit. Annals of the New York Academy of Sciences 655: 349-364.
	
\endbiblist
}

\end{document}